\documentclass{article}

\usepackage[final]{neurips_2022}

\usepackage{amsmath}
\usepackage{graphicx}
\usepackage{multirow}
\usepackage[linesnumbered,ruled,vlined]{algorithm2e}

\usepackage[flushleft]{threeparttable}

\usepackage[utf8]{inputenc} 
\usepackage[T1]{fontenc}    
\usepackage{hyperref}       
\usepackage{url}            
\usepackage{booktabs}       
\usepackage{amsfonts}       
\usepackage{nicefrac}       
\usepackage{microtype}      
\usepackage{xcolor}         

\title{Triadic Temporal Exponential Random Graph Models}

\author{%
  Yifan Huang \\
  Modeling and Simulation Department \\
  University of Central Florida\\
  Orlando, FL, U.S.A \\
  \And
  Clayton Barham\\
  Department of Computer Science\\
  University of Central Florida\\
  Orlando, FL, U.S.A \\
  \AND
  Eric Page \\
  Department of Electrical and Computer Engineering \\
  University of Central Florida \\
  Orlando, FL, U.S.A \\
  \And
  PK Douglas \\
  IACS, Santa Monica, CA, U.S.A.\\
  Department of Psychiatry and Biobehavioral Medicine, UCLA\\
  Los Angeles, CA, U.S.A
}

\begin{document}

\maketitle

\begin{abstract}
Temporal exponential random graph models (TERGM) are powerful statistical models that can be used to infer the temporal pattern of edge formation and elimination in complex networks (e.g., social networks). TERGMs can also be used in a generative capacity to predict longitudinal time series data in these evolving graphs. However, parameter estimation within this framework fails to capture many real-world properties of social networks, including: triadic relationships, small world characteristics, and social learning theories which could be used to constrain the probabilistic estimation of dyadic covariates. Here, we propose triadic temporal exponential random graph models (TTERGM) to fill this void, which includes these hierarchical network relationships within the graph model. We represent social network learning theory as an additional probability distribution that optimizes Markov chains in the graph vector space. The new parameters are then approximated via Monte Carlo maximum likelihood estimation. We show that our TTERGM model achieves improved fidelity and more accurate predictions compared to several benchmark methods on GitHub network data.
\end{abstract}

\section{Introduction}
Social networks have been studied for over a century, and graph theory techniques have been used for decades to model relationship patterns amongst individuals and social entities \citep{wasserman1994social}.  Graph network statistics have provided fundamental and theoretical insight into social phenomenal across political, economic and behavioral data.  However, historically, these techniques have been applied to study static ‘snapshots’ of a network, limiting the ability to make large-scale inferences about social network dynamics.

In recent years, online social networks have provided a wealth of open source empirical data, providing new opportunities to test and adjudicate amongst competing theories and quantitative models \citep{borgatti2009network}. Online social networks are formed by nodes (e.g., individuals) and edges, embedding social relationships (e.g., friends, relatives, colleagues) within a complex graph network. Both network structure and nodal connectivity constraint the flow of information or resources through the network \citep{kane2014s}. However, online social networks are ever expanding, and their connectivity evolves rapidly over time. Thus, statistically modeling these data for generative and predictive purposes is challenging. 

Temporal exponential random graph models (TERGM) have become a core tool for modeling dynamic social networks. Although a variety of approaches including variations of TERGM have been applied to education \citep{mamas2020friendship}, finance, \citep{park2018structural} and political data \citep{2016Election}, few TERGM variations have attempted to quantify the effect of "influencers" alongside triadic relationships within a dynamic network. Here, we expand the classic TERGM model to support triadic relationships to make predictions on dynamic graphs.  We apply this new triadic temporal exponential random graph model (TTERGM) to data derived from Github. Github has been studied extensively within the context of graphical network modeling due to its popularity, open access, and transparency. Thus, these data provide a benchmark for model comparison. GitHub also provides network programming functionality \citep{borges2016understanding}, offering the opportunity to study "influencers", their affect on network structure, and triadic relationships within a dynamic network. 

\section{Related Work}
Social network models have been applied for many purposes to include: modeling an individual's behavioral patterns to predict future nodal attributes (e.g., connections) over time \citep{mcconnell2018everybody} \citep{mcavoy2020social}, modeling  interactions and cluster formation within  online communities \citep{fortunato2016community} \citep{xu2020industry} \citep{liu2018competition}, and modeling how network characteristics (e.g., centrality) influences its users \citep{qiu2017hidden}. Overall, these models attempt to characterize the relationship amongst network structure and information diffusion, decision making, and individual behavior. \citep{jackson2017economic}.

General classes of network formation methods include: 1) exponential random graph models (ERGMs) \citep{lusher2013exponential}\citep{pattison1999logit}, meta-networks, and meta-matrices \citep{carley2001structural}\citep{krackhardt1998pcans} for multilayer social networks. 2) block modeling \citep{guimera2009missing} 3) geographic or characteristic based approaches,  \citep{boucher2012my}\citep{leung2014random}; 4) link formation techniques \citep{christakis2010empirical}\citep{bramoulle2012homophily}, and 5) subgraph model-based approaches (SUGMs) \citep{chandrasekhar2016network}. Numerous studies have examined the formation of cascades of network activity, characterizing and predicting network growth \citep{bakshy2012role}\citep{gruhl2004information}\citep{yang2010predicting}. Typically, spikes of activity occur within a few days of content’s introduction into the network. This property forms the backdrop to a line of temporal analyses that focus on the basic rising-and-falling pattern that characterizes the initial occurrence of information burst.

Influence on the Github platform can be quantified by the number of followers, stars, mentions, quotes, and up-votes received from other users. Social network metrics such as centrality indicate how broadly influence extends (e.g. geographic interest) \citep{weber2014makes}. Other features include project size, file volume, critical folder, lines of code and calling of basic functions. The popularity rate can be measured by (Total\_Stars / project\_life). Few studies have examined influence of user-popularity, repo-popularity, and triadic relationships in dynamic graphs.

\section{Methodology}
\subsection{Preliminaries}
Exponential random graph models (ERGM) are static. Temporal exponential random graph models (TERGM) are an extension of ERGMs to handle dynamic information in real-world online social networks \citep{hanneke2010discrete}. ERGM can be written as in \eqref{ERGM2}. TERGM with Markov assumption can be written as in \eqref{TERGM}. 

\begin{equation}
\label{ERGM2}
P(N) = \frac{1}{Z}exp(\sum{\gamma(N))}
\end{equation}

\begin{equation}
\label{TERGM}
P(N^t | N^{t-1}) = \frac{1}{Z_{t-1}}exp(\sum_{ij}{\sigma(N_{ij}^t,N_{ij}^{t-1}))}
\end{equation}

\begin{figure}
\includegraphics[width=140mm]{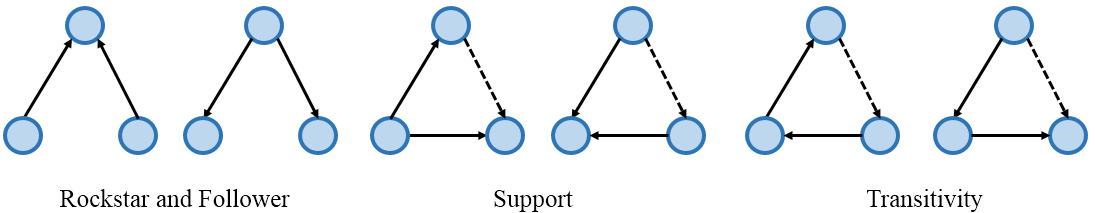}
\centering
\caption{Examples of Triadic relationship in Influencer Networks}
\label{example}
\end{figure}

$P(N)$ represents the probability of the network $N$; $Z$ represents the normalization constant that is usually difficult to compute; $\gamma$ is the vector of network characteristics such as number of edges, triangles, 2-stars, etc. $t$ represents the sequence of network observations; $\sigma$ is the vector of social-theory driven temporal network characteristics such as homophily, transitivity, reciprocity, etc. $t$ represents the sequence of network observations; Compared to ERGMs, TERGMs are able to model the distribution on time series data (either embedded in the network or in separate timesteps), hence certain temporal patterns can be captured and reflected in the parameter values for phenomenon interpretation. Examples of these dynamic patterns of triadic effects in influencer networks are shown in Figure \ref{example}. TERGM provides significant advantages over ERGMs, since certain static patterns can be enriched in higher dimensions when the sequence order is considered. TERGMs are also capable of modeling observed friendship networks with bootstrap methods estimated by maximum pseudolikelihood \citep{leifeld2018temporal}, or networks of infectious disease transmission using statistical methods in network analysis \citep{jenness2018epimodel}. The flexibility of TERGMs make it possible to adapt to a variety of input data types, such as cross-sectional or longitude data \citep{henry2016analyzing}\citep{block2018change}. 

\subsection{Triadic Temporal Exponential Random Graph Model}
\label{model_description}
\begin{figure}
\includegraphics[width=140mm]{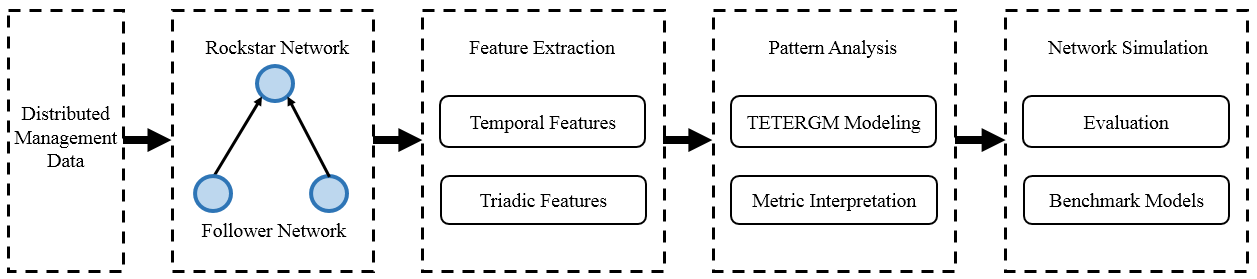}
\centering
\caption{Design of the Triadic Temporal Exponential Random Graph Model}
\label{TETERGM}
\end{figure}

 TERGM models estimated within the markov chain assumption are typically incapable of generating and reproducing realistic dynamics observed in real-world online social networks. We hypothesized that increasing the model's capacity to describe triadic network properties would reduce the error between the model and empirical observations. We propose TTERGM here to sequentially predict network probabilities by integrating the dynamics between influencers and followers. TTERGM was run on a computer with 12900K CPU, 1080TI and 128GB RAM. Figure \ref{TETERGM} shows the framework of TTERGM that has five major components - data collection module, network processing module, feature extraction module, pattern analysis module, and a generative network module. 

Influencer-follower networks $N_t$ were identified and constructed by connecting users by events in Table \ref{EventCategory}. In the subsequent module, temporal features and triadic features listed in Table \ref{RockstarMetrics} were extracted from the influencer-follower networks. Network characteristics were then estimated using the Markov chain Monte Carlo (MCMC) method. In the pattern analysis module, the TTERGM model was applied to model the data. The general form of TTERGM can be written as in Equation \eqref{TTERGM}. $P(N)$ represents the probability of a given network architechture, N; $Z$ represents the normalization constant as done in classic TERGM models; $\gamma$ is a vector of network characteristics (e.g, number of edges, triangles, 2-stars), $t$ represents the temporal sequence of network observations, and $\sigma$ is the vector of social-theory driven temporal network characteristics such as homophily, transitivity, reciprocity, etc. To fit the TTERGM model to the data, algorithm \ref{algorithm} is used to initialize and traverse each node in the network to construct the generative influencer-follower networks. Finally, we used these observed features to simulate real-world influencer-follower networks and compared the predictive of TTERGM performance with the classic TERGM model and the Block Model using left-out validation data.

\begin{equation}
\label{TTERGM}
P(N^t | N^{t-1}) = \frac{1}{Z_{t-1}}exp(\sum_{ij}{\sigma(N_{ij}^t,N_{ij}^{t-1})) + ... + \frac{1}{Z_{1}}exp(\sum_{ij}{\sigma(N_{ij}^2,N_{ij}^{1}))} } 
\end{equation}

\begin{algorithm}[H]
 \large
 \KwIn{Online social network $N$}
 Initialization $G = G_{t_0}$\
 \Repeat{k iterations}{
 \While{$ t \neq t_n$}{
  read current node $n_{i}$\;
  \eIf{$n_{i}$ is not traversed}{
    \For{each link $l_{ij}$ in graph $G_{t_0}$}{
        calculate the network statistics of each link and \\
        fill  the values to the feature vector $N_{ij}$ \\
        $Prob(l_{ij}) =  \sum_{p=0,q=0}^{pq}{Prob(l_{ij} | l_{pq})}$ \\
    }
  }{
   read the next node $n_{i+1}$\;
  }
  $t = t+1$\;
 }
 Calculate $P(N^t) = \frac{1}{Z}exp(\sum_{ij}{\sigma(N_{ij}^t,N_{ij}^{t-1},... ,N_{ij}^{t_0})}$ where \\
 $Z = MCMC(exp(\sum_{ij}{u(N_{ij}^t,N_{ij}^{t-1},... ,N_{ij}^{t_0})})$ where\\
 $Z$ is the normalization constant, $u$ is the network statistics,
 $MCMC$ is the Markov Chain Monte Carlo method \\
 }
 \caption{Triadic Temporal Exponential Random Graph for Network Generation}
 \label{algorithm}
\end{algorithm}

\section{Experiments and Results}
\label{Results}

The GitHub dataset is from 1/1/2015 to 8/30/2017 including 2 million users and 13 million projects. We selected the 100 most popular repositories because we aimed to characterize top GitHub repositories, and the impact of influencers on the popularity of repositories using the TTERGM model. The number of repositories is a hyperparameter which can be adjusted, depending on the goal of the model. We selected the top 10 users, in terms of number of followers, as the influencers in this study. This threshold was chosen because the number of followers drops off sharply after that point, but can be chosen arbitrarily for different datasets. The data was acquired using the API from GitHub \citep{gousios2012ghtorrent}. The API can be used to stream GitHub repository interactions with customized formats. Optionally, meta data from user relation events can be retrieved as well. The dataset contains 14 types of events which are listed in Table \ref{EventCategory}.

\begin{table}[ht!]
\begin{center}
\caption{Event Categories and Descriptions}
\label{EventCategory}
\begin{tabular}{p{3cm} p{5cm} p{5cm}} 
  \textbf{Event Category} & \textbf{Event Type} & \textbf{Description}\\
  \hline
  \multirow{1}{3cm}{Receptive Events} & WatchEvent  & When someone stars a repository\\
  & PullRequestReviewCommentEvent & When comment on a pull request’s unified changes\\
  & IssueCommentEvent & When an issue comment is created or edited\\
  & MemberEvent & When a user is invited or removed as a collaborator to a repository\\
  & IssuesEvent & When an issue is created or edited\\
  & GollumEvent & When a wiki page is created or edited\\
  \hline
  \multirow{1}{3cm}{Contributive Events} & WatchEvent  & When someone stars a repository\\
  & ForkEvent & When an user forks a repository\\
  & ReleaseEvent & When a release is published or edited\\
  & PublicEvent & When a private repository is made public\\
  & PullRequestEvent & When a pull request is created or edited\\
  & PushEvent & When a push is happened to a repository branch\\  
  & DeleteEvent & When a branch or tag is deleted\\  
  & CommitCommentEvent & When is commit comment is created\\  
  & CreateEvent & When a branch or tag is created\\  
\end{tabular}
\end{center}
\end{table}

We implemented the TTERGM technique to model the dynamics between repository popularity and networks. The 3 most followed anonymized influencers had 52,722, 30,161, and 25,827 followers respectively. Each of the top 10 most popular influencer has at least 14 thousand followers. We divided the 15 types of events into 2 categories - participative events and contributive events to highlight the social characteristics. Participative events demonstrate a user's engagement to a project repository. Contributive events indicate the cooperation among developers in a software repository. Followers tend to have more participative events than influencers, while influencers generally have more contributive events than followers.

\begin{table}[ht!]
\begin{center}
\caption{Features Extracted from Influencer Network}
\label{RockstarMetrics}
\begin{tabular}{p{3cm} p{4cm} p{5cm}} 
  \textbf{Measurement Type} & \textbf{Measurement Name} & \textbf{Description}\\
  \hline
  \multirow{1}{3cm}{Network connection} & \# direct links  & Number of direct links influencers have with their followers\\
  &\# indirect length-2 links & Number of indirect links of length 2 from influencers to their followers\\
  &\# indirect length-3 links & Number of indirect links of length 3 from influencers to their followers\\
  &\# triangles & Influencer’s activity on a repository triggered his/her follower’s activity on the same repository, then a triangle forms\\  
  \hline
  \multirow{1}{3cm}{Network topology} & Average shortest path & Average shortest path length of pairs of nodes in the network\\
  & Assortativity & Pearson correlation coefficient of degree between pairs of linked nodes\\
  &\# of connected components & Number of connected components in the network\\
  & Average Clustering Coefficient & Ratio of number of triangles over maximum possible number of triangles\\    
  &\# nodes & Number of nodes in the network\\    
  &\# edges & Number of edges in the network\\    
\end{tabular}
\end{center}
\end{table}

\begin{table}[ht!]
\begin{center}
\caption{Number of Followers of the Top 10 Influencers}
\label{RockstarFollowers}
\begin{tabular}{p{2cm}|p{6cm}|p{2cm}} 
  \textbf{Rank} & \textbf{Anonymized Influencer Id} & \textbf{Number of Followers}\\
  \hline
  1 & lBMOoXAjxIN\_Dc3alQNLZQ & 52722\\
  2 & BhQS5KA8AvmQJXbsVeusdw & 30161\\
  3 & s0jAeLRt2onrivaUCqdJrg & 25827\\
  4 & QFB1aZ8GXkNYHyfWe7aEeA & 24604\\
  5 & jAGnWUFUmnBc9ydeQbIfDQ & 24510\\
  6 & hXalEIoEWnEbCSfiQI1LNA & 23076\\
  7 & eUnkVgArKJiNOBhb0w53\_Q & 18522\\
  8 & VRyyOPSJUCS5jRlDtwjefA & 15755\\
  9 & wNDkYd6NACSuvLCnxog23w & 15396\\
  10 & wHfAzUFXU8D186qTl9c54w & 14928\\
\end{tabular}
\end{center}
\end{table}

\begin{table}[ht!]
\begin{center}
\caption{Block, TERGM, and TTERGM models were used to generate
predicted distributions for network characteristics on out-of-sample temporal observations}
\label{hypothesis2}
\begin{tabular}{c p{2.5cm} p{2.5cm} p{2.5cm} p{2.5cm} p{0.1cm}} 
  \hline
  Network Model & 2017-07 in-deg & 2017-07 out-deg & 2017-08 in-deg & 2017-08 out-deg \\
  \hline
  \multicolumn{6}{l}{\hspace{4cm}   \hspace{10cm}   }\\
  Block Model & 4.39 & 5.08 & 5.32 & 4.56 \\
  \multicolumn{6}{l}{\hspace{4cm}   \hspace{10cm}   }\\
  TERGM & 4.65 & 4.78 & 5.13 & 4.35\\
  \multicolumn{6}{l}{\hspace{4cm}   \hspace{10cm}   }\\
  TTERGM & 3.42\textsuperscript{*} & 4.15\textsuperscript{*} & 4.25\textsuperscript{*} & 3.25\textsuperscript{*}\\
  \end{tabular}

  \begin{tablenotes}
    \small
    {\raggedright  Note: $*$ indicates p-value < 0.05 comparing to Block Model and TERGM respectively \par}
    \end{tablenotes}   
  
\end{center}
\end{table}

To evaluate the performance of the TTERGM, we compared the simulation with two benchmark models - TERGM and Block model. Each model is evaluated using a set of metrics (average degree of incoming edges and average degree of outgoing edges) to see how well the predicted network distribution matched the real-world network characteristics. The evaluation result (average of 30 runs) was shown in Table \ref{hypothesis2}. The Block model and TERGM perform similarly in month 2017-07 and 2017-08. Comparing to TERGM, TTERGM has 26.45\% less errors in 2017-07 for incoming degree of errors, 13.17\% less errors in 2017-07 for outgoing degree of errors, 17.15\% less errors in 2017-08 incoming degree of errors, and 25.28\% less errors in 2017-08 for outgoing degree of errors. We believe the consistent improvement stems from the extra computation from TTERGM in the markov chain.   

\section{Conclusion}
\label{Conclusion}
We implemented a social-theory driven temporal exponential random graph model to infer the temporal pattern of edge formation and elimination in complex networks (e.g., social networks), and examine the effect of influencers and triadic relatinships on predicting future network dynamics. When popular repositories are formed or influencers act, the structure of the social network alters, affecting network metrics. The TTERGM technique build upon previous statistical models by incorporating information flow across hierarchical configuration features. We represent social network learning theory as an additional probability distribution that optimizes Markov chains in the graph vector space. The new parameters are then approximated via Monte Carlo maximum likelihood estimation. The TTERGM model is capable of reproducing the dynamics observed empirically in large-scale social network data, and produced more accurate predictions on left-out data compared to the classic TERGM and block models. However, the TTERGM model imposes additional computational burden during parameter estimation, which may hinder its ability to scale to larger datasets.  Future work may include expanding this approach to model the influence of more "distant" users in the network, or those that do not directly follow an influencer.

\section{Acknowledgements}
\label{Acknowledgements}
We are very thankful for support from the Air Force Office of Scientific Research under award number FA9550-20-1-0042, which helped make this work possible.

\bibliography{fin}

\end{document}